\documentclass{svmult}
\usepackage{epsfig,graphicx} 
\usepackage{natbib,url}      
\bibpunct{(}{)}{;}{a}{}{,}   
\def\cite#1{\citealp{#1}}    
\def\authorindex#1{}  

\begin{document}\newcount\preprintheader\preprintheader=1



\def\thisvolume{these proceedings}

\def\aj{{AJ}}			
\def\araa{{ARA\&A}}		
\def\apj{{ApJ}}			
\def\apjl{{ApJ}}		
\def\apjs{{ApJS}}		
\def\ao{{Appl.\ Optics}} 
\def\apss{{Ap\&SS}}		
\def\aap{{A\&A}}		
\def\aapr{{A\&A~Rev.}}		
\def\aaps{{A\&AS}}		
\def\an{{Astron.\ Nachrichten}}
\def\aspcs{{ASP Conf.\ Ser.}}
\def\assp{{Astrophys.\ \& Space Sci.\ Procs., Springer, Heidelberg}}
\def\azh{{AZh}}			
\def\baas{{BAAS}}		
\def\jrasc{{JRASC}}	
\def\memras{{MmRAS}}		
\def\mnras{{MNRAS}}
\def\nat{{Nat}}		
\def\pra{{Phys.\ Rev.\ A}} 
\def\prb{{Phys.\ Rev.\ B}}		
\def\prc{{Phys.\ Rev.\ C}}		
\def\prd{{Phys.\ Rev.\ D}}		
\def\prl{{Phys.\ Rev.\ Lett.}} 
\def\pasp{{PASP}}
\def\pasj{{PASJ}}		
\def\qjras{{QJRAS}}
\def\science{{Sci}}		
\def\skytel{{S\&T}}		
\def\solphys{{Solar\ Phys.}} 
\def\sovast{{Soviet\ Ast.}}  
\def\ssr{{Space\ Sci.\ Rev.}}
\def\svassp{{Astrophys.\ Space Sci.\ Procs., Springer, Heidelberg}}
\def\zap{{ZAp}}			
\let\astap=\aap
\let\apjlett=\apjl
\let\apjsupp=\apjs
\def\grl{{Geophys.\ Res.\ Lett.}}  
\def\jgr{{J. Geophys.\ Res.}} 

\def\ion#1#2{{\rm #1}\,{\uppercase{#2}}}  
\def\deg{\hbox{$^\circ$}}
\def\sun{\hbox{$\odot$}}
\def\earth{\hbox{$\oplus$}}
\def\la{\mathrel{\hbox{\rlap{\hbox{\lower4pt\hbox{$\sim$}}}\hbox{$<$}}}}
\def\ga{\mathrel{\hbox{\rlap{\hbox{\lower4pt\hbox{$\sim$}}}\hbox{$>$}}}}
\def\sq{\hbox{\rlap{$\sqcap$}$\sqcup$}}
\def\arcmin{\hbox{$^\prime$}}
\def\arcsec{\hbox{$^{\prime\prime}$}}
\def\fd{\hbox{$.\!\!^{\rm d}$}}
\def\fh{\hbox{$.\!\!^{\rm h}$}}
\def\fm{\hbox{$.\!\!^{\rm m}$}}
\def\fs{\hbox{$.\!\!^{\rm s}$}}
\def\fdg{\hbox{$.\!\!^\circ$}}
\def\farcm{\hbox{$.\mkern-4mu^\prime$}}
\def\farcs{\hbox{$.\!\!^{\prime\prime}$}}
\def\fp{\hbox{$.\!\!^{\scriptscriptstyle\rm p}$}}
\def\micron{\hbox{$\mu$m}}
\def\onehalf{\hbox{$\,^1\!/_2$}}	
\def\onethird{\hbox{$\,^1\!/_3$}}
\def\twothirds{\hbox{$\,^2\!/_3$}}
\def\onequarter{\hbox{$\,^1\!/_4$}}
\def\threequarters{\hbox{$\,^3\!/_4$}}
\def\ubv{\hbox{$U\!BV$}}		
\def\ubvr{\hbox{$U\!BV\!R$}}		
\def\ubvri{\hbox{$U\!BV\!RI$}}		
\def\ubvrij{\hbox{$U\!BV\!RI\!J$}}		
\def\ubvrijh{\hbox{$U\!BV\!RI\!J\!H$}}		
\def\ubvrijhk{\hbox{$U\!BV\!RI\!J\!H\!K$}}		
\def\ub{\hbox{$U\!-\!B$}}		
\def\bv{\hbox{$B\!-\!V$}}		
\def\vr{\hbox{$V\!-\!R$}}		
\def\ur{\hbox{$U\!-\!R$}}


\def\labelitemi{{\bf --}}  

\def\rmit#1{{\it #1}}              
\def\rmit#1{{\rm #1}}              
\def\etal{\rmit{et al.}}           
\def\etc{\rmit{etc.}}           
\def\ie{\rmit{i.e.,}}              
\def\eg{\rmit{e.g.,}}              
\def\cf{cf.}                       
\def\viz{\rmit{viz.}}
\def\vs{\rmit{vs.}}

\def\rot{\hbox{\rm rot}}
\def\div{\hbox{\rm div}}
\def\lesssim{\mathrel{\hbox{\rlap{\hbox{\lower4pt\hbox{$\sim$}}}\hbox{$<$}}}}
\def\gtrsim{\mathrel{\hbox{\rlap{\hbox{\lower4pt\hbox{$\sim$}}}\hbox{$>$}}}}
\def\mathstacksym#1#2#3#4#5{\def#1{\mathrel{\hbox to 0pt{\lower 
    #5\hbox{#3}\hss} \raise #4\hbox{#2}}}}
\mathstacksym\lesssim{$<$}{$\sim$}{1.5pt}{3.5pt} 
\mathstacksym\gtrsim{$>$}{$\sim$}{1.5pt}{3.5pt} 
\mathstacksym\lrarrow{$\leftarrow$}{$\rightarrow$}{2pt}{1pt} 
\mathstacksym\lessgreat{$>$}{$<$}{3pt}{3pt} 

\def\dif{\: {\rm d}}                       
\def\ep{\:{\rm e}^}                        
\def\dash{\hbox{$\,-\,$}}                  
\def\is{\!=\!}                             

\def\starname#1#2{${#1}$\,{\rm {#2}}}  
\def\Teff{\hbox{$T_{\rm eff}$}}   

\def\kms{\hbox{km$\;$s$^{-1}$}}
\def\ms{\hbox{m$\;$s$^{-1}$}}
\def\Mxcm{\hbox{Mx\,cm$^{-2}$}}    

\def\Bapp{\hbox{$B_{\rm app}$}}    

\def\komega{($k, \omega$)}                 
\def\kf{($k_h,f$)}                         
\def\VminI{\hbox{$V\!\!-\!\!I$}}           
\def\IminI{\hbox{$I\!\!-\!\!I$}}           
\def\VminV{\hbox{$V\!\!-\!\!V$}}           
\def\Xt{\hbox{$X\!\!-\!t$}}                

\def\level #1 #2#3#4{$#1 \: ^{#2} \mbox{#3} ^{#4}$}   

\def\specchar#1{\uppercase{#1}}    
\def\AlI{\mbox{Al\,\specchar{i}}}  
\def\BI{\mbox{B\,\specchar{i}}} 
\def\BII{\mbox{B\,\specchar{ii}}}  
\def\BaI{\mbox{Ba\,\specchar{i}}}  
\def\BaII{\mbox{Ba\,\specchar{ii}}} 
\def\CI{\mbox{C\,\specchar{i}}} 
\def\CII{\mbox{C\,\specchar{ii}}} 
\def\CIII{\mbox{C\,\specchar{iii}}} 
\def\CIV{\mbox{C\,\specchar{iv}}} 
\def\CaI{\mbox{Ca\,\specchar{i}}} 
\def\CaII{\mbox{Ca\,\specchar{ii}}} 
\def\CaIII{\mbox{Ca\,\specchar{iii}}} 
\def\CoI{\mbox{Co\,\specchar{i}}} 
\def\CrI{\mbox{Cr\,\specchar{i}}} 
\def\CriI{\mbox{Cr\,\specchar{ii}}} 
\def\CsI{\mbox{Cs\,\specchar{i}}} 
\def\CsII{\mbox{Cs\,\specchar{ii}}} 
\def\CuI{\mbox{Cu\,\specchar{i}}} 
\def\FeI{\mbox{Fe\,\specchar{i}}} 
\def\FeII{\mbox{Fe\,\specchar{ii}}} 
\def\FeIX{\mbox{Fe\,\specchar{ix}}}
\def\FeX{\mbox{Fe\,\specchar{x}}}
\def\FeXVI{\mbox{Fe\,\specchar{xvi}}}
\def\FrI{\mbox{Fr\,\specchar{i}}}
\def\HI{\mbox{H\,\specchar{i}}} 
\def\HII{\mbox{H\,\specchar{ii}}} 
\def\Hmin{\hbox{\rmH$^{^{_{\scriptstyle -}}}$}}      
\def\Hemin{\hbox{{\rm He}$^{^{_{\scriptstyle -}}}$}} 
\def\HeI{\mbox{He\,\specchar{i}}} 
\def\HeII{\mbox{He\,\specchar{ii}}} 
\def\HeIII{\mbox{He\,\specchar{iii}}} 
\def\KI{\mbox{K\,\specchar{i}}} 
\def\KII{\mbox{K\,\specchar{ii}}} 
\def\KIII{\mbox{K\,\specchar{iii}}} 
\def\LiI{\mbox{Li\,\specchar{i}}} 
\def\LiII{\mbox{Li\,\specchar{ii}}} 
\def\LiIII{\mbox{Li\,\specchar{iii}}} 
\def\MgI{\mbox{Mg\,\specchar{i}}} 
\def\MgII{\mbox{Mg\,\specchar{ii}}} 
\def\MgIII{\mbox{Mg\,\specchar{iii}}} 
\def\MnI{\mbox{Mn\,\specchar{i}}} 
\def\NI{\mbox{N\,\specchar{i}}}
\def\NIV{\mbox{N\,\specchar{iv}}}
\def\NaI{\mbox{Na\,\specchar{i}}}
\def\NaII{\mbox{Na\,\specchar{ii}}}
\def\NaIII{\mbox{Na\,\specchar{iii}}}
\def\NeVIII{\mbox{Ne\,\specchar{viii}}} 
\def\NiI{\mbox{Ni\,\specchar{i}}} 
\def\NiII{\mbox{Ni\,\specchar{ii}}}
\def\NiIII{\mbox{Ni\,\specchar{iii}}} 
\def\OI{\mbox{O\,\specchar{i}}} 
\def\OVI{\mbox{O\,\specchar{vi}}}
\def\RbI{\mbox{Rb\,\specchar{i}}} 
\def\SII{\mbox{S\,\specchar{ii}}} 
\def\SiI{\mbox{Si\,\specchar{i}}} 
\def\SiII{\mbox{Si\,\specchar{ii}}} 
\def\SrI{\mbox{Sr\,\specchar{i}}}
\def\SrII{\mbox{Sr\,\specchar{ii}}}
\def\TiI{\mbox{Ti\,\specchar{i}}} 
\def\TiII{\mbox{Ti\,\specchar{ii}}} 
\def\TiIII{\mbox{Ti\,\specchar{iii}}} 
\def\TiIV{\mbox{Ti\,\specchar{iv}}} 
\def\VI{\mbox{V\,\specchar{i}}} 
\def\HtwoO{\mbox{H$_2$O}}        
\def\Otwo{\mbox{O$_2$}}          

\def\Halpha{\mbox{H\hspace{0.1ex}$\alpha$}} 
\def\Ha{\mbox{H\hspace{0.2ex}$\alpha$}}
\def\Hbeta{\mbox{H\hspace{0.2ex}$\beta$}}
\def\Hgamma{\mbox{H\hspace{0.2ex}$\gamma$}}
\def\Hdelta{\mbox{H\hspace{0.2ex}$\delta$}}
\def\Hepsilon{\mbox{H\hspace{0.2ex}$\epsilon$}}
\def\Hzeta{\mbox{H\hspace{0.2ex}$\zeta$}}
\def\Lyalpha{\mbox{Ly$\hspace{0.2ex}\alpha$}}
\def\Lybeta{\mbox{Ly$\hspace{0.2ex}\beta$}}
\def\Lygamma{\mbox{Ly$\hspace{0.2ex}\gamma$}}
\def\Lycont{\mbox{Ly\hspace{0.2ex}{\small cont}}}
\def\Baalpha{\mbox{Ba$\hspace{0.2ex}\alpha$}}
\def\Babeta{\mbox{Ba$\hspace{0.2ex}\beta$}}
\def\Bacont{\mbox{Ba\hspace{0.2ex}{\small cont}}}
\def\Paalpha{\mbox{Pa$\hspace{0.2ex}\alpha$}}
\def\Bralpha{\mbox{Br$\hspace{0.2ex}\alpha$}}

\def\NaD{\mbox{Na\,\specchar{i}\,D}}    
\def\NaDone{\mbox{Na\,\specchar{i}\,\,D$_1$}}
\def\NaDtwo{\mbox{Na\,\specchar{i}\,\,D$_2$}}
\def\NaID{\mbox{Na\,\specchar{i}\,\,D}}
\def\NaIDone{\mbox{Na\,\specchar{i}\,\,D$_1$}}
\def\NaIDtwo{\mbox{Na\,\specchar{i}\,\,D$_2$}}
\def\Done{\mbox{D$_1$}}
\def\Dtwo{\mbox{D$_2$}}

\def\Mgbone{\mbox{Mg\,\specchar{i}\,b$_1$}}
\def\Mgbtwo{\mbox{Mg\,\specchar{i}\,b$_2$}}
\def\Mgbthree{\mbox{Mg\,\specchar{i}\,b$_3$}}
\def\MgIb{\mbox{Mg\,\specchar{i}\,b}}
\def\MgIbone{\mbox{Mg\,\specchar{i}\,b$_1$}}
\def\MgIbtwo{\mbox{Mg\,\specchar{i}\,b$_2$}}
\def\MgIbthree{\mbox{Mg\,\specchar{i}\,b$_3$}}

\def\CaIIK{\mbox{Ca\,\specchar{ii}\,K}}       
\def\CaIIH{\mbox{Ca\,\specchar{ii}\,H}}
\def\CaIIHK{\mbox{Ca\,\specchar{ii}\,H\,\&\,K}}
\def\HK{\mbox{H\,\&\,K}}
\def\Kthree{\mbox{K$_3$}}      
\def\Hthree{\mbox{H$_3$}}
\def\Ktwo{\mbox{K$_2$}}
\def\Htwo{\mbox{H$_2$}}
\def\Kone{\mbox{K$_1$}}     
\def\Hone{\mbox{H$_1$}}     
\def\KtwoV{\mbox{K$_{2V}$}}
\def\KtwoR{\mbox{K$_{2R}$}}
\def\KoneV{\mbox{K$_{1V}$}}
\def\KoneR{\mbox{K$_{1R}$}}
\def\HtwoV{\mbox{H$_{2V}$}}
\def\HtwoR{\mbox{H$_{2R}$}}
\def\HoneV{\mbox{H$_{1V}$}}
\def\HoneR{\mbox{H$_{1R}$}}

\def\hk{\mbox{h\,\&\,k}}
\def\kthree{\mbox{k$_3$}}    
\def\hthree{\mbox{h$_3$}}
\def\ktwo{\mbox{k$_2$}}
\def\htwo{\mbox{h$_2$}}
\def\kone{\mbox{k$_1$}}     
\def\hone{\mbox{h$_1$}}     
\def\ktwoV{\mbox{k$_{2V}$}}
\def\ktwoR{\mbox{k$_{2R}$}}
\def\koneV{\mbox{k$_{1V}$}}
\def\koneR{\mbox{k$_{1R}$}}
\def\htwoV{\mbox{h$_{2V}$}}
\def\htwoR{\mbox{h$_{2R}$}}
\def\honeV{\mbox{h$_{1V}$}}
\def\honeR{\mbox{h$_{1R}$}}

\ifnum\preprintheader=1     
\makeatletter  
\def\@maketitle{\newpage
\markboth{}{}%
  {\mbox{} \vspace*{-8ex} \par 
   \em \footnotesize To appear in ``Magnetic Coupling between the Interior 
       and the Atmosphere of the Sun'', eds. S.~S.~Hasan and R.~J.~Rutten, 
       Astrophysics and Space Science Proceedings, Springer-Verlag, 
       Heidelberg, Berlin, 2009.} \vspace*{-5ex} \par
 \def\lastand{\ifnum\value{@inst}=2\relax
                 \unskip{} \andname\
              \else
                 \unskip \lastandname\
              \fi}%
 \def\and{\stepcounter{@auth}\relax
          \ifnum\value{@auth}=\value{@inst}%
             \lastand
          \else
             \unskip,
          \fi}%
  \raggedright
 {\Large \bfseries\boldmath
  \pretolerance=10000
  \let\\=\newline
  \raggedright
  \hyphenpenalty \@M
  \interlinepenalty \@M
  \if@numart
     \chap@hangfrom{}
  \else
     \chap@hangfrom{\thechapter\thechapterend\hskip\betweenumberspace}
  \fi
  \ignorespaces
  \@title \par}\vskip .8cm
\if!\@subtitle!\else {\large \bfseries\boldmath
  \vskip -.65cm
  \pretolerance=10000
  \@subtitle \par}\vskip .8cm\fi
 \setbox0=\vbox{\setcounter{@auth}{1}\def\and{\stepcounter{@auth}}%
 \def\thanks##1{}\@author}%
 \global\value{@inst}=\value{@auth}%
 \global\value{auco}=\value{@auth}%
 \setcounter{@auth}{1}%
{\lineskip .5em
\noindent\ignorespaces
\@author\vskip.35cm}
 {\small\institutename\par}
 \ifdim\pagetotal>157\p@
     \vskip 11\p@
 \else
     \@tempdima=168\p@\advance\@tempdima by-\pagetotal
     \vskip\@tempdima
 \fi
}
\makeatother     
\fi


\title*{Outstanding Issues in Solar Dynamo Theory}


\author{D. Nandy}


\authorindex{Nandy, D.}


\institute{Indian Institute of Science Education and Research, Kolkata,
           India}

\maketitle

\setcounter{footnote}{0}  

\begin{abstract}

The magnetic activity of the Sun, as manifested in the sunspot cycle,
originates deep within its convection zone through a dynamo mechanism
which involves non-trivial interactions between the plasma and
magnetic field in the solar interior. Recent advances in
magnetohydrodynamic dynamo theory have led us closer towards a better
understanding of the physics of the solar magnetic cycle. In
conjunction, helioseismic observations of large-scale flows in the
solar interior has now made it possible to constrain some of the
parameters used in models of the solar cycle. In the first part of
this review, I briefly describe this current state of understanding of
the solar cycle. In the second part, I highlight some of the
outstanding issues in solar dynamo theory related to the the nature of
the dynamo $\alpha$-effect, magnetic buoyancy and the origin of
Maunder-like minima in activity. I also discuss how poor
constraints on key physical processes such as turbulent diffusion,
meridional circulation and turbulent flux pumping confuse the relative
roles of these vis-a-vis magnetic flux transport. I argue that
unless some of these issues are addressed, no model of the solar cycle
can claim to be ``the standard model'', nor can any predictions from
such models be trusted; in other words, we are still not there yet.

\end{abstract}

\section{Introduction}

Sunspots have been telescopically observed for centuries, starting with the pioneering
observations of many, including Galileo Galilei in the early $17^{th}$ century. Much later in the
$20^{th}$ century, Schwabe discovered that the number of sunspots on the solar surface vary
cyclically, and Carrington discovered that the sunspots appear at lower and lower solar latitudes
with the progress of the cycle. With Hale's discovery of magnetic fields within sunspots in 1908,
it became clear that the sunspot cycle is in fact a magnetic cycle (see Figure~1 for an overview
of the magnetic butterfly diagram). Efforts to theoretically explain the origin of the solar
cycle continued from then on and took a giant leap in 1955 when Parker outlined his theory of the
solar cycle based on a magnetohydrodynamic (MHD) dynamo mechanism.

In what follows, I briefly summarize the important concepts underlying
the solar cycle that have been developed in the last half of the
$20^{th}$ century (Section~2). I then describe the current state of
our understanding, concentrating on those ideas that are widely
accepted as important for dynamo action (Section~3). Following this, I
highlight the outstanding issues that need to be addressed towards
developing a ``standard model'' of the solar cycle (Section~4).
Finally, I end with some concluding remarks (Section~5).

Before we proceed, it is important here to state the scope of this review; this is neither meant
to be a comprehensive review of all complementary ideas in solar dynamo theory and modeling, nor
is it a reference source for important works in this field. Interested readers who desire these,
are referred to the recent, and comprehensive review on the solar dynamo by Charbonneau
(2005)\nocite{charbonneau05}. This is a personalized account of the field as I perceive it to be.

\begin{figure}[t]
\centerline{\includegraphics[width=11cm]{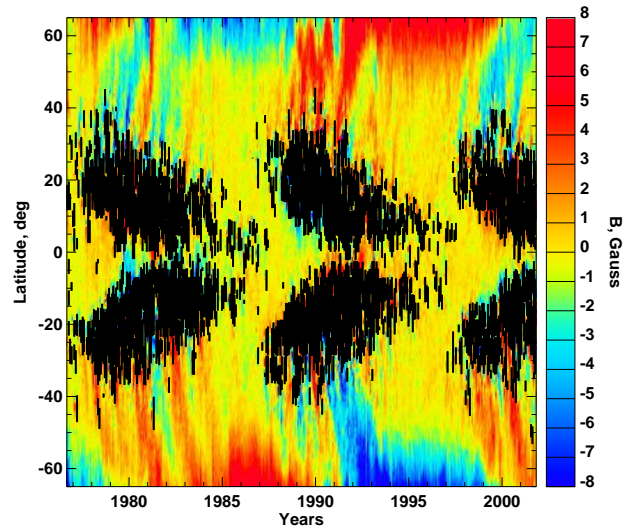}}
\caption{The solar butterfly diagram depicting the latitude of sunspot appearance (think dark
lines) with time. The background shows the weak and diffuse field outside of sunspots. Note that
while the sunspot formation belt migrates equatorward, the weak field outside of it migrates
poleward with the progress of the cycle, reversing the older polar field at the time of sunspot
maximum.}
\end{figure}

\section{Basic concepts}

The interior of the Sun consists of highly ionized gas, \ie\ plasma. The fundamental equation
which governs the behavior (and generation) of magnetic fields in such a plasma system is the
induction equation
\begin{equation}
\frac{\partial \bf{B}}{\partial t} =  {\nabla} \times (\bf{v} \times \bf{B} - \eta \, \nabla
\times \bf{B)},
\end{equation}
where {\bf{B}} is the magnetic field, {\bf{v}} the velocity field and $\eta$ the effective
magnetic diffusivity of the system. In astrophysical systems such as the Sun, the plasma has a
very high characteristic magnetic Reynolds number (the ratio of the first to the second term on
the R.H.S.\ of the above equation). In such a plasma the magnetic fields are frozen in the fluid
and therefore the field and plasma movement are coupled. This allows the energy of convective
flows in the solar convection zone (SCZ) to be drawn into producing and amplifying magnetic
fields, which is the essence of the dynamo mechanism.

Under the approximation of spherical symmetry, applicable to a star such as the Sun, the magnetic
and velocity fields can be expressed as
\begin{equation}
{\bf B} = B_{\phi} {\bf \hat{e}}_{\phi} + \nabla \times (A {\bf \hat{e}}_{\phi})
\end{equation}
\begin{equation}
  {\bf v} = r\sin(\theta)\Omega{\bf \hat{e}}_{\phi} + {\bf v}_p.
\end{equation}
The first term on the R.H.S.\ of Equation~$2$ is the toroidal component (i.e, in the
$\phi$-direction) and the second term is the poloidal component (i.e., in the $r$-$\theta$ plane)
of the magnetic field. In the case of the velocity field (Equation~3), these two terms correspond
to the differential rotation $\Omega$ and meridional circulation $v_p$, respectively. The field
of helioseismology has now constrained the profile of the solar differential rotation throughout
the solar convection zone and it is therefore no longer a free parameter in models of the solar
dynamo. The meridional circulation is observed in the surface and helioseismic inversions
constrain it somewhat in the upper $10\%$ of the Sun, however, the deeper counter-flow is not yet
observed and is theoretically constructed by invoking mass conservation in conjunction with the
solar density stratification.

Since the Sun rotates differentially, any pre-existing poloidal field would get stretched in the
direction of rotation creating a toroidal component. Such horizontal toroidal flux tubes in the
solar interior are subject to magnetic buoyancy (Parker 1955a\nocite{parker55a}) and therefore
erupt out through the surface creating bipolar sunspot pairs. These bipolar sunspot pairs acquire
a tilt due to the action of Coriolis force during their rise through the solar convection zone
(SCZ), generating what is commonly known as the Joy's law distribution of solar active region
tilt angles. To complete the dynamo chain of events, the toroidal component of the magnetic field
has to be converted back into the poloidal component. This necessitates the action of a
non-axisymmetric mechanism, i.e., with non-zero vorticity. The first such proposed mechanism was
due to Parker (1955b)\nocite{parker55b} who proposed that small-scale helical turbulence can
twist rising toroidal flux tubes back into the $r$-$\theta$ plane thereby recreating the poloidal
field -- a process which traditionally came to be known as the dynamo $\alpha$-effect; this
became an essential ingredient in models of the solar dynamo.

\section{Current state of our understanding}

In the last two decades, simulations of the dynamics of buoyantly rising (thin) toroidal flux
tubes showed that the initial strength of these flux tubes at the base of the SCZ had to be on
the order of $10^5$ G, to match the morphological properties of active regions observed at the
solar surface (D'Silva \& Choudhuri 1993\nocite{dsilva-choudhuri93}; Fan, Fisher \& DeLuca
1994\nocite{fan-fisher-deluca93}). However, the equipartition field strength of the magnetic
field in the SCZ (i.e., the field strength at which magnetic and convective flow energies are in
equipartition) is of the order of $10^4$ G. If the strength of the sunspot forming toroidal flux
tubes are an order of magnitude greater than the equipartition field strength, then the helical
convective flows would be unable to twist them as envisaged in the traditional dynamo
$\alpha$-effect formalism.

This realization has now led the dynamo community to explore alternative mechanisms for the
regeneration of the poloidal field. Amongst the various contenders, the so-called
Babcock-Leighton (BL) mechanism is perceived to be the front-runner. In this mechanism,
originally proposed by Babcock (1961)\nocite{babcock61} and Leighton 
(1969)\nocite{leighton69}, the decay of tilted bipolar sunspot pairs,
and the subsequent (net) poleward dispersal of their flux by surface
processes such as diffusion, differential rotation and meridional
circulation regenerates and reverses the solar poloidal field (Dikpati
\& Charbonneau 1999\nocite{dikpati-charbonneau99}; Nandy \& Choudhuri
2001\nocite{nandy-choudhuri01}). Although this process is now commonly
referred to as the BL $\alpha$-effect, it may be noted that in spirit,
this process is very different from the traditional $\alpha$-effect;
in the latter formalism, averaging over small scale turbulence and the
first-order-smoothing-approximation is required, whereas in the
former, it is not. The BL mechanism for poloidal field generation is
actually observed on the surface and has been substantiated with
numerical surface-flux-transport simulations.

Since this poloidal field generation mechanism is primarily located in the near-surface layers,
the generated poloidal field has to be transported back into the solar interior, where the
toroidal field amplification and storage takes place (in the overshoot layer at the base of the
SCZ). In most BL models of the solar cycle, this flux transport is achieved by meridional
circulation, although turbulent diffusion is also expected to play a significant role. In such
models, it is found that the meridional circulation governs the spatio-temporal distribution of
sunspots on the solar surface (Nandy \& Choudhuri 2002\nocite{nandy-choudhuri02}) and its speed
determines the period of the solar cycle, even in regimes where the SCZ is diffusion dominated
(Yeates, Nandy \& Mackay 2008\nocite{yeates08}).

The amplitude of the sunspot cycle is found to be weakly correlated with the speed of the
meridional circulation and the coefficient of turbulent diffusion. However, the threshold of
magnetic buoyancy, i.e., the field strength at which stored toroidal flux tubes become
magnetically buoyant and escape out of the overshoot layer, is a limiting factor on the amplitude
of the solar cycle (Nandy 2002\nocite{nandy02}).

\section{Outstanding issues}

Although it seems that we have made much progress in the last decade or so in understanding many
aspects of the solar cycle, this progress has also uncovered multiple aspects which pose a
challenge to dynamo theory. I take this opportunity to discuss some of these outstanding issues.

\subsection{Nature of the dynamo $\alpha$-effect}

While the BL mechanism for poloidal field regeneration is observed on the solar surface, a few
other $\alpha$-effect mechanisms have been proposed in recent times, which do not have any direct
observational confirmation, but nevertheless may be functional in the solar interior. These
$\alpha$-effects are driven by magnetic field instabilities or differential rotation
instabilities and are spatially located around the base of the SCZ (for an overview of the
various proposed dynamo $\alpha$-effects, see Charbonneau 2005\nocite{charbonneau05}).

What is unclear is the extent to which these proposed $\alpha$-effects may contribute to poloidal
field regeneration in the Sun and the relative efficacy of these compared to the BL mechanism;
this is connected to the following question (I believe first articulated by Manfred Sch\"ussler):
Is the observed BL mechanism a by-product of a dynamo mechanism that completely resides in the
solar interior, or is it actually an integral part the dynamo mechanism? Although the success of
dynamo models based on the BL mechanism argues for the latter scenario, in my view we still
cannot rule out the possibility that other $\alpha$-effect mechanisms may contribute at least in
parts to poloidal field generation. If this were to be the case it creates an interesting dilemma
which is  explained below.

For some time now many of us are content with the perception that we can observe one complete
half of the solar dynamo mechanism, namely the poloidal field regeneration at the solar surface
through active regions decay and dispersal. These surface observations have been used widely to
constrain and fine tune dynamo models. Moreover, because the poloidal field of a given cycle
feeds directly into producing the toroidal field of future cycle(s), these surface observations
provide an useful tool to predict future cycle amplitudes. However, if some other,
observationally unconstrained mechanism for poloidal field regeneration is actually more dominant
than the BL mechanism, this would pose a serious challenge to our current perceptions of the
solar cycle and would negatively impact attempts to predict future solar activity. Therefore, any
evidence related to mechanisms of solar poloidal field generation has to be seriously evaluated
to illuminate whether there are multiple mechanisms and if yes, what are their relative
contributions to the overall dynamo.

\subsection{Treatment of magnetic buoyancy}

An issue coupled to the nature of the poloidal field regeneration mechanism is the treatment
magnetic buoyancy and bipolar sunspot creation in models of the solar cycle. It is believed that
those strong toroidal flux tubes stored in the stably stratified region beneath the SCZ, which
exceed a certain threshold (on the order of $100$ kG), become magnetically buoyant when they
emerge out in the SCZ and subsequently produce poloidal field. This whole process of buoyant
eruption and poloidal field generation is treated in dynamo models through diverse
implementations -- almost all of which are not fully consistent with the philosophy of the BL
mechanism.

The most popular approach, in the context of the BL dynamo, has been to approximate this process
with a poloidal field source term that is located at near-surface layers (constrained by a
prescribed spatially dependent function) and which is proportional to the toroidal field strength
at the base of the SCZ. Although this approach typically has a upper quenching threshold which
stops poloidal field creation when the toroidal field exceeds a certain threshold (in accordance
with flux tube rise simulations which show that very strong toroidal flux tubes come out without
any tilt), most modelers do not use a lower operational threshold. This causes even weak, sub-kG
toroidal field to contribute to poloidal field creation through the BL mechanism, which goes
against the spirit of the BL idea. An alternative approach, which has been used by some modelers,
is to employ an explicit algorithm for magnetic buoyancy which searches for strong toroidal
fields exceeding the buoyancy threshold and transporting this field to the surface layers,
conserving flux in the process. However, this process too over-simplifies the
surface-flux-dispersal process as it still uses a source term at the surface. Basically, the
usage of this source-term preempts the surface-flux-transport process, which results in dynamo
simulations giving results that are not consistent with surface-flux-transport simulations when a
variable meridional flow is used (Schrijver \& Liu 2008\nocite{schrijver08}).

A more realistic, but rarely used approach is to buoyantly erupt spaced double rings of opposite
radial field (akin to bipolar sunspot pairs) to the solar surface using a explicit buoyancy
algorithm (for a comparative study of various buoyancy algorithms, including the double-ring
approach, see Nandy \& Choudhuri 2001\nocite{nandy-choudhuri01}). The source-term is discarded
with in this approach. Surface differential rotation, meridional circulation and diffusion
subsequently acts on these erupted double rings to generate the poloidal field in a truer
representation of the BL philosophy. However, this approach is computationally intensive as it
demands a very high grid-resolution, which is sufficiently close to solar AR spatial-scale. This
being a impractical task, the over-simplified and somewhat questionable buoyancy prescriptions
continue to be used in solar dynamo models. If an alternative, physically correct algorithm
cannot be devised, it seems the brute-force solution to this problem is to implement more
computationally efficient numerical algorithms for the solar dynamo that can handle very high
grid-resolutions.

\subsection{Origin of grand minima}

Small, but significant variations in solar cycle amplitude is commonly observed from one cycle to
another and models based on either stochastic fluctuations, or non-linear feedback, or time-delay
dynamics exist to explain such variability in cycle amplitude (for overviews see, Charbonneau
2005\nocite{charbonneau05}; Wilmot-Smith et al.\ 2006\nocite{wilmot-smith06}). However, most
models find it difficult to switch off the sunspot cycle completely for an extended period of
time -- such as that observed during the Maunder minimum -- and subsequently recover back to
normal activity.

Two important and unresolved questions in this context are what physical mechanism stops active
region creation completely and how does the dynamo recover from this quiescent state. The first
question is the more vexing one and still eludes a coherent and widely accepted explanation. The
second question is less challenging in my opinion; the answer possibly lies in the continuing
presence of another $\alpha$-effect (could be the traditional dynamo $\alpha$-effect suggested by
Parker) which can work on weaker, sub-equipartition toroidal fields -- to slowly build up the
dynamo amplitude to eventually recover the sunspot cycle from a Maunder-like grand minima.

These are speculative ideas and one thing that can be said with confidence at this writing is
that we are just scratching the surface as far as the physics of grand minima like episodes is
concerned.

\subsection{Parametrization of turbulent diffusivity}

Typically, in many dynamo models published in the literature, the coefficient of turbulent
diffusivity employed in much lower than that suggested by mixing-length theory (about $10^{13}$
cm$^2$/s; Christensen-Dalsgaard et al.\ 1996\nocite{Christensen-Dalsgaardetal96}). This is done
to ensure that the flux transport in the SCZ in advection dominated (i.e., meridional circulation
is the primary flux transport process). There are many disadvantages to using a higher
diffusivity value in these dynamo models. Usage of higher diffusivity values makes the flux
transport process diffusion dominated, reducing the dynamo period to values somewhat lower than
the observed solar cycle period. It also makes flux storage and amplification difficult and
shortens cycle memory; the latter is the basis for solar cycle predictions. Nevertheless, this
inconsistency between mixing-length theory and parametrization of turbulent diffusivity in dynamo
models is, in my opinion, a vexing problem.

In the absence of any observational constraints on the depth-dependence of the diffusivity
profile in the solar interior, this problem can only be addressed theoretically. One possible
solution to resolving this inconsistency is by invoking magnetic quenching of the mixing-length
theory suggested diffusivity profile. The idea is simple enough; since magnetic fields have an
inhibiting effect on turbulent convection, strong magnetic fields should quench and thereby be
subject to less diffusive mixing. The magnetic quenching of turbulent diffusivity is challenging
to implement numerically, but seems to me to be the best bet towards reconciling this
inconsistency within the framework of the current modeling approach.

\subsection{Role of downward flux pumping}

An important physical mechanism for magnetic flux transport has been identified recently from
full MHD simulations of the solar interior. This mechanism, often referred to as turbulent flux
pumping, pumps magnetic field preferentially downwards, in the presence of rotating, stratified
convection such as that in the SCZ (see e.g., Tobias et al.\ 2001\nocite{tobias01}). Typical
estimates yield a downward pumping speed which can be as high as $10$ m/s; this would make flux
pumping the dominant downward flux transport mechanism in the SCZ, short--circuiting the
transport by meridional circulation and turbulent diffusion. However, turbulent flux pumping is
usually ignored in kinematic dynamo models of the solar cycle.

If indeed the downward pumping speed is as high as indicated, then turbulent flux pumping may
influence the solar cycle period, crucially impact flux storage and amplification and also affect
solar cycle memory. Therefore, turbulent flux pumping must be properly accounted for in kinematic
dynamo models and its effects completely explored; this remains an issue to be addressed
adequately.

\section{Concluding remarks}

Now let us elaborate on, and examine some of the consequences of the
outstanding issues highlighted in the earlier section.

\subsection{A story of communication timescales}

To put a broader perspective on some of these issues facing dynamo
theory, specifically in the context of the interplay between various
flux-transport processes, it will be instructive here to consider the
various timescales involved within the dynamo mechanism. Let us, for
the sake of argument, consider that the BL mechanism is the
predominant mechanism for poloidal field regeneration. Because this
poloidal field generation happens at surface layers, but toroidal
field is stored and amplified deeper down near the base of the SCZ,
for the dynamo to work these two spatially segregated layers must
communicate with each other. In this context, magnetic buoyancy plays
an important role in transporting toroidal field from the base of the
SCZ to the surface layers -- where the poloidal field is produced. The
timescale of buoyant transport is quite short, on the order of $0.1$
year and this process dominates the upward transport of toroidal
field.

Now, to complete the dynamo chain, the poloidal field must be brought back down to deeper layers
of the SCZ where the toroidal field is produced and stored. There are multiple processes that
compete for this downward transport, namely meridional circulation, diffusion and turbulent flux
pumping.

Considering the typical meridional flow loop from mid-latitudes at the
surface to mid-latitudes at the base of the SCZ, and a peak flow speed
of $20$ m/s, one gets a typical circulation timescale $\tau_v$ = 10
years. Most modelers use low values of diffusivity on the order of
$10^{11}$ cm$^2$/s, which makes the diffusivity timescale (${L_{\rm
SCZ}^2}/\eta$, assuming vertical transport over the depth of the SCZ),
$\tau_{\eta}$ = 140 years; i.e., much more that $\tau_v$, therefore
making the circulation dominate the flux transport. However, if one
assumes diffusivity values close to that suggested by mixing length
theory (say, $5$ $\times$ $10^{12}$ cm$^2$/s), then the diffusivity
timescale becomes $\tau_{\eta}$ = 2.8 years; i.e., shorter than the
circulation timescale -- making diffusive dispersal dominate the flux
transport process.

If we now consider the usually ignored process of turbulent pumping, the situation changes again.
Assuming a typical turbulent pumping speed on the order of $10$ m/s over the depth of the SCZ
gives a timescale $\tau_{\rm pumping}$ = 0.67 years, shorter than both the diffusion and meridional
flow timescales. This would make turbulent pumping the most dominant flux transport mechanism for
downward transport of poloidal field into the layers where the toroidal field is produced and
stored.

\subsection{Solar cycle predictions}

As outlined in Yeates, Nandy \& Mackay (2008)\nocite{yeates08}, the length of solar cycle memory
(defined as over how many cycles, the poloidal field of a given cycle would contribute to
toroidal field generation) determines the input for predicting the strength of future solar
cycles. The relative timescales of different flux transport mechanisms within the dynamo chain of
events, and their interplay -- based on which process (or processes) dominate, determine this
memory. For example, if the dynamo is advection (circulation) dominated then the memory tends to
be long, lasting over multiple cycles. However, if the dynamo is diffusion (or turbulent pumping)
dominated, then this memory would be much shorter.

Now, within the scope of the current framework of dynamo models, I have argued that significant
confusion exists regarding the role of various flux transport processes. So much so that we do
not yet have a consensus on which of these processes dominate; therefore we do not have a
so-called ``standard-model'' of the solar cycle yet. Should solar cycle predictions be trusted
then?

Taking into account this uncertainty in the current state of our understanding of the solar
dynamo mechanism, I believe that any solar cycle predictions -- that does not adequately address
these outstanding issues -- should be carefully evaluated. In fact, under the circumstances, it
is fair to say that if any solar cycle predictions match reality, it would be more fortuitous
than a vindication of the model used for the prediction. This is not to say that modelers should
not explore the physical processes that contribute to solar cycle predictability; indeed that is
where most of our efforts should be. My concern is that we do not yet understand all the physical
processes that constitute the dynamo mechanism and their interplay well enough to begin making
predictions. Prediction is the ultimate test of any model, but there are many issues that need to
be sorted out before the current day dynamo models are ready for that ultimate test.

\begin{acknowledgement}

This work has been supported by the Ramanujan Fellowship of the Department of Science and
Technology, Government of India and a NASA Living With a Star Grant NNX08AW53G to the Smithsonian
Astrophysical Observatory at Harvard University. I gratefully acknowledge many useful
interactions with colleagues at the solar physics groups at Montana State University (Bozeman)
and the Harvard Smithsonian Center for Astrophysics (Boston). I am indebted to my friends at
Bozeman, Montana, from where I recently moved back to India, for contributing to a very enriching
experience during the seven years I spent there.

\end{acknowledgement}

\begin{small}

\end{small}

\end{document}